# Study of infrared scintillations in gaseous and liquid argon – Part II: light yield and possible applications

**A. Bondar,**[a,b] **A. Buzulutskov,**[a,b,*] **A. Dolgov,**[b] **A. Grebenuk,**[a] **S. Peleganchuk,**[a,b] **V. Porosev,**[a] **L. Shekhtman,**[a,b] **E. Shemyakina**[a,b] **and A. Sokolov**[a,b]

[a] *Budker Institute of Nuclear Physics SB RAS, Lavrentiev avenue 11, 630090 Novosibirsk, Russia*
[b] *Novosibirsk State University, Pirogov street 2, 630090 Novosibirsk, Russia*
*E-mail: A.F.Buzulutskov@inp.nsk.su*

ABSTRACT: We present here a comprehensive study of the light yield of primary and secondary scintillations produced in gaseous and liquid Ar in the near infrared (NIR) and visible region, at cryogenic temperatures. The measurements were performed using Geiger-mode avalanche photodiodes (GAPDs) and pulsed X-ray irradiation. The primary scintillation yield of the fast emission component in gaseous Ar was found to be independent of temperature in the range of 87-160 K; it amounted to 17000±3000 photon/MeV in the NIR in the range of 690-1000 nm. In liquid Ar at 87 K, the primary scintillation yield of the fast component was considerably reduced, amounting to 510±90 photon/MeV, in the range of 400-1000 nm. Proportional NIR scintillations (electroluminescence) in gaseous Ar were also observed; their amplification parameter at 160 K was measured to be 13 photons per drifting electron per kV. No proportional scintillations were observed in liquid Ar up to the electric fields of 30 kV/cm. The applications of NIR scintillations in dark matter search and coherent neutrino-nucleus scattering experiments and in ion beam radiotherapy are considered.

KEYWORDS: Near infrared scintillations in noble gases and liquids; Electroluminescence; Rare-event experiments.

---

[*] Corresponding author.

## Contents

## 1. Introduction

Following the first results obtained in our laboratory [1], we present here a comprehensive study of the light yield of primary and secondary scintillations produced in gaseous and liquid Ar in the near infrared (NIR) and visible region at cryogenic temperatures. The study of primary and secondary scintillations in noble gases and liquids is of paramount importance to rare-event experiments using noble gas media: see for example reviews [2],[3],[4]. In particular in two-phase noble-gas detectors [3],[4],[5], the comparison of primary scintillations in the noble liquid to those of secondary in the gas phase helps to select WIMP signals against the background in dark matter search experiments [6],[7],[8]. Here secondary scintillations mean proportional scintillations, i.e. those of electroluminescence generated by electrons drifting under a moderate electric field in the gas phase [4],[3]. The other examples of noble gas scintillation applications are the noble liquid scintillation calorimetry of the μ-e-γ decay search experiment [9] based on recording primary scintillations in liquid Xe, the high-pressure noble-gas TPC of the double-β decay search experiment [10] based on recording proportional scintillations in compressed Xe, and the two-phase Ar and Xe detectors for coherent neutrino-nucleus scattering experiments [11],[12] based on recording proportional scintillations in the gas phase.

In another class of detectors based on secondary scintillations, avalanche scintillations are employed for optical readout, i.e. electroluminescence under higher electric fields at which the avalanche multiplication occurs in the noble gas medium. This technique consists of optically recording avalanche-induced scintillations emitted from the holes of GEMs [13] or THGEMs [14]: at room temperature - using CCD cameras [15] and large area APDs [16]; at cryogenic temperatures - using CCD cameras [17],[18] and Geiger-mode APDs (GAPDs or SiPMs, [19]) [20],[21],[22],[23].

The latter technique is applied in so-called Cryogenic Avalanche Detectors (CRADs, [2]) with optical readout using combined THGEM/GAPD multiplier, operated in two-phase Ar or Xe. Here the optical readout is performed in either the Vacuum Ultraviolet (VUV) [20],[23] or NIR [21],[22] region, using Wavelength Shifter (WLS)-coated or uncoated GAPDs respectively. Such CRADs are aimed at reaching single electron sensitivity at low noise, i.e. at having ultimately low detection threshold for primary ionization. Accordingly, they may come to be in great demand in rare-event experiments with low energy deposition such as those of coherent neutrino-nucleus scattering and dark matter search [2].

It should be remarked that the present study has actually been triggered by the recent results on NIR scintillations obtained in our laboratory in the course of CRAD developments [21],[22]: a rather high avalanche (secondary) scintillation yield, of about 4 NIR photons per avalanche (secondary) electron, was observed in the two-phase Ar CRAD with THGEM/GAPD optical readout. Note that such a high scintillation yield was obtained in the NIR, which is not a trivial fact.

Indeed, until recently noble gas scintillations in high energy physics experiments have been studied essentially in the vacuum ultraviolet (VUV): see for example [3],[24],[25],[26]. This necessitated the use of sophisticated VUV-sensitive photodetectors. In the VUV the primary scintillation yield is rather high: of about $(40-60)*10^3$ photon/MeV in liquid Ar and Xe [3] and $14*10^3$ photon/MeV in gaseous Xe [24]. The VUV emission is caused by reactions between excited and ionized atoms producing excimers which decay radiating the VUV continua [27]. At high pressures this emission was generally believed to dominate over all other

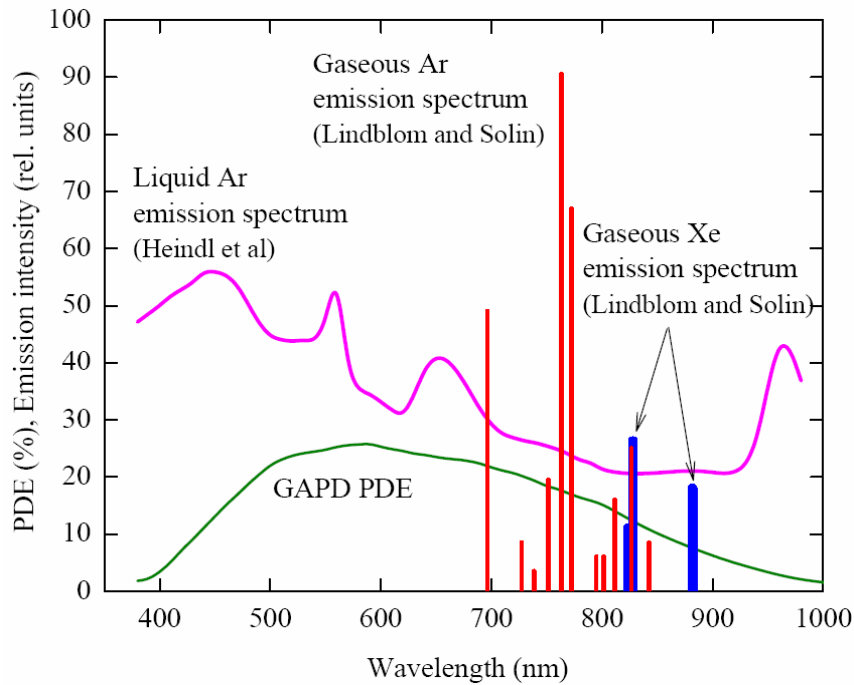

Fig. 1. Emission spectra in the visible and NIR region in gaseous Ar and Xe due to proton beam-induced scintillations [28] and in liquid Ar due to electron beam-induced scintillations [31],[32]. The Photon Detection Efficiency (PDE) of the GAPD "CPTA 149-35" is also shown [37],[38].

types of radiative decays such as atomic emission in the visible and infrared regions [3].

On the other hand, as early as 20 years ago it was recognized that this statement might not be valid due to the discovery of intense atomic emission scintillations in practically all noble gases in the non-VUV region, namely in the NIR [28], in particular in Ar in the wavelength range of 690-850 nm [28] and in Xe at 800-1600 nm [29]: see Fig. 1 showing the appropriate emission spectra. Since then, the NIR emission spectra of scintillations in gaseous and liquid Ar and Xe have been further studied in several works: both for primary [29],[30],[31],[32] and secondary [30],[33],[34] scintillations. In gaseous Ar this kind of scintillation was attributed to transitions between the atomic states of the Ar ($3p^5 4p$) and Ar ($3p^5 4s$) configurations [28],[33]. In contrast, the non-VUV emission spectrum of liquid Ar is continuous, extending from 400 to 1000 nm [31],[32]: see Fig. 1; its emission mechanism has not been yet clarified. However, little was known about the absolute NIR scintillation yield in noble gases: practically nothing about that in Ar and only the lower limit in gaseous Xe ($\geq 21*10^3$ photon/MeV) [34],[35]. For more information see ref. [2], where the state of the art in the field of NIR scintillations and their applications for CRAD developments has been reviewed.

First results on NIR scintillation yield in gaseous and liquid Ar were presented in a short letter [1]. Here we present a more elaborated paper on that subject. The present article is preceded by Part I [36], dedicated to the general description of a novel methodology to measure NIR scintillations, namely to that using pulsed X-ray irradiation and NIR-sensitive GAPDs at cryogenic temperatures in a single photoelectron counting mode with time resolution. In addition, the time structure of NIR scintillations was studied in Part I [36]. In the present paper (Part II), the experimental results on the light yield of primary and secondary NIR scintillations are presented, as well as possible applications of NIR scintillations in rare-event experiments and ion beam radiotherapy.

## 2. Primary and secondary scintillation yield: terminology and definitions

In the following sections, we will use the notions of primary, secondary, proportional and avalanche scintillations, as well as that of electroluminescence. To avoid confusion, let us settle their terminology and definitions of their light yields.

Primary scintillation is defined as that produced by primary ionization. In particular at zero electric field, only primary scintillation can be produced by ionizing radiation. Secondary scintillation or electroluminescence is defined as that produced by the electrons (primary or secondary) drifting in an electric field. In noble gases at moderate electric fields, such secondary scintillation is sometimes called proportional scintillation, since its intensity is proportional to the electric field in a certain field range.

Proportional scintillations in noble gases are caused by atomic excitation processes. At higher fields the latter are taken over by atomic ionization processes, i.e. by those of electron avalanche multiplication. Accordingly, at higher electric fields proportional scintillations are taken over by avalanche scintillations, the intensity of the latter being not any more proportional to the electric field.

Let us now consider the definitions of the scintillation (light) yields used in the present study.

In part I [36], the time structure of NIR scintillations was studied: fast and slow emission components were observed, with time constants below 0.1 μs and of about 20-30 μs respectively, both in gaseous and liquid Ar [36]. While the fast component is expected from the atomic emission mechanism of NIR scintillations, due to fast transitions between the atomic states of the Ar $(3p^5\,4p)$ and Ar $(3p^5\,4s)$ configurations [28],[33],[39], the origin of the slow component has not been yet understood. Accordingly in the following, the scintillation yield in gaseous and liquid Ar will be presented for the fast component only.

In addition in Part I [36], a new methodology to measure NIR scintillation yield in gaseous and liquid Ar at cryogenic temperatures was developed. It consists of the irradiation of a given parallel-plate gap (filled with Ar) with pulsed X-rays and measuring the number of electrons ($N_e$) and that of scintillation photons emitted over full solid angle ($N_{ph}$), produced in the gap per X-ray pulse. To verify the validity of the measurement procedure, the measurements were conducted in two gaps, "top" and "bottom", viewed by two different GAPDs, top and bottom respectively: see Fig. 1 in ref. [36]. Accordingly, the scintillation (light) yield measured in a given gap is as follows:

$$Y = \frac{N_{ph}}{N_e} = \frac{N_{pe}}{<PDE>A_{GAPD}\,N_e} \quad . \tag{1}$$

This formula is valid both for primary and secondary scintillations. Here the number of photons was measured with the help of the GAPD, counting the number of photoelectrons ($N_{pe}$) contained in a scintillation signal, taking into account the GAPD Photon Detection Efficiency averaged over the Ar emission spectrum ($<PDE>$) and the GAPD acceptance with respect to the X-ray conversion region in the gap ($A_{GAPD}$, equivalent to the average reduced solid angle $<\Delta\Omega/4\pi>$).

For secondary scintillations produced in a parallel-plate gap in a proportional scintillation mode, it is convenient to define the electroluminescence (light) yield as the number of emitted photons ($N_{ph}$), the primary scintillation contribution being subtracted, normalized to the drifting electron and to the unit drift path, i.e. divided by the total ionization charge generated in the gap ($N_e$) and by the average electron drift path in the gap ($l$):

$$Y_{el} = \frac{N_{ph}}{N_e\,l} \quad . \tag{2}$$

In our case, the primary ionization in gaseous Ar was produced practically uniformly across the gap. Accordingly, in a proportional scintillation mode when the gap is operated in the ionization mode (i.e. without secondary ionization generated in the gap), the average electron drift path is equal to the half of the gap thickness ($d$):

$$l = d/2 \quad . \tag{3}$$

When the parallel-plate gap starts to operate in a proportional ionization mode (i.e. with charge avalanche multiplication), the average electron drift path in the gap should be corrected since it becomes to be dependent on the charge gain ($G$), due to avalanche electrons tending to be produced closer to the anode (see Appendix for formula derivation):

$$l = \frac{d}{2}\,\frac{G-1}{G\ln G} \quad . \tag{4}$$

To compare the data obtained at different pressures and temperatures, it is generally accepted to parameterize the reduced electroluminescence yield ($Y_{el}/N$) as a linear function of the reduced electric field ($E/N$), where $N$ is the atomic density (see for example [25]), for the proportional scintillation regime:

$$\frac{Y_{el}}{N}\,[10^{-17}\,\text{photon electron}^{-1}\,\text{cm}^2\,\text{atom}^{-1}] = a\,\frac{E}{N} - b \quad . \tag{5}$$

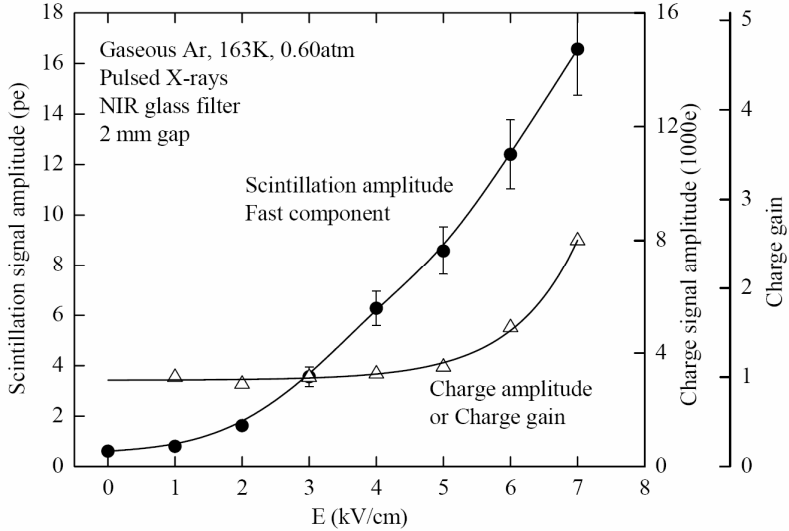

Fig. 2. Typical scintillation and charge signals amplitudes for the fast emission component, measured in the top gap in gaseous Ar at 163 K and 0.60 atm, with NIR glass filter in front of the GAPD. Shown are the scintillation signal amplitude of the fast component expressed in photoelectrons (left scale) and the charge signal amplitude expressed in electrons (right scale), per X-ray pulse, as a function of the electric field in the gap. The charge gain (right scale) is also shown.

Here $E/N$ is given in Td [$10^{-17}$ V cm$^2$ atom$^{-1}$]. This equation is universally valid for any temperature and pressure. Parameters $a$ and $b$ are determined by the particular scintillation mechanism: $a$ defines the slope of the field dependence which characterizes the electroluminescence efficiency, while $b$ defines the threshold at which electroluminescence starts. It follows from this equation that far above the threshold, the electroluminescence intensity at a given temperature is independent of pressure. Moreover it is fully determined by the product of the electric field and the gap thickness, i.e. by the voltage applied across the gap, in accordance to the hypothesis that the energy obtained by the drifting electron from the electric field is fully converted to proportional scintillations in the VUV and NIR regions.

## 3. Primary scintillation yield in gaseous Ar

In this paper we present the results obtained at cryogenic temperatures: in gaseous Ar - at temperatures of about 160 K and 123 K, at a gas density corresponding to that of 1 atm at room temperature, and at 87 K and 1.0 atm, i.e. in saturated vapour; in liquid Ar - at 87 K and 1.0 atm. The oxygen-equivalent impurity content in Ar was below 20 ppb, while the N$_2$ content was below 10 ppm [36]. The scintillations in the GAPD sensitivity region, i.e. in the NIR and visible region, were observed under these conditions and their light yields were measured in gaseous and liquid Ar, under irradiation with pulsed X-rays with the energy of 15-40 keV and 30-40 keV respectively.

Fig. 2 illustrates typical behavior of scintillation and charge signal amplitudes as a function of the electric field, i.e. $N_{pe}$ and $N_e$ respectively, in gaseous Ar as measured in the top gap. One can see that in gaseous Ar, at low and moderate electric fields applied across the gap, varying from 0.1 to 5 kV/cm, the gap was operated in an ionization mode. i.e. without electron multiplication. At these fields the collected charge was practically independent of the field, enabling to unambiguously determine the primary ionization charge ($N_e$), including at zero electric field using extrapolation procedure (see Fig.2). On the other hand, the scintillation signal starts increasing much earlier, at 2 kV/cm, indicating upon the appearance of secondary (proportional) scintillations in the NIR.

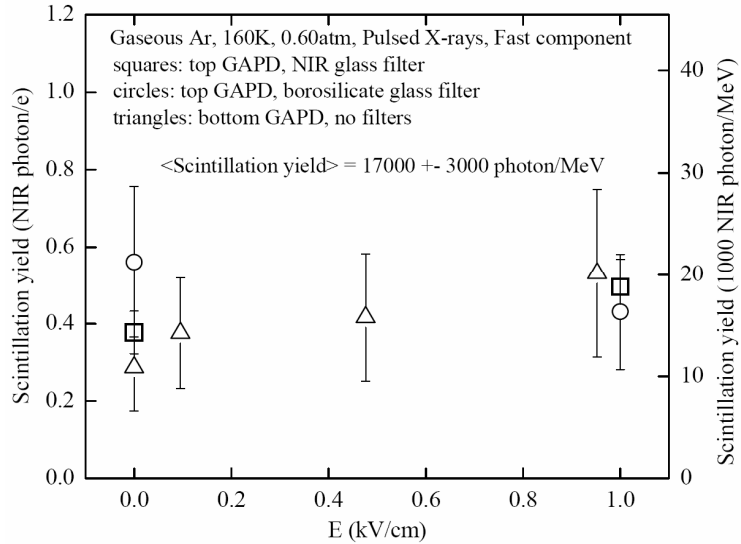

Fig. 3. Primary scintillation yield in the NIR for the fast emission component in gaseous Ar at 160 K and 0.60 atm, measured in the electric field range of 0-1 kV/cm under different conditions: with the top GAPD and NIR glass filter, with the top GAPD and borosilicate glass filter, with the bottom GAPD and no optical filter. The scintillation yield is expressed in the number of photons over $4\pi$ per primary ionization electron (left scale) and per MeV of deposited energy of the primary ionization (right scale).

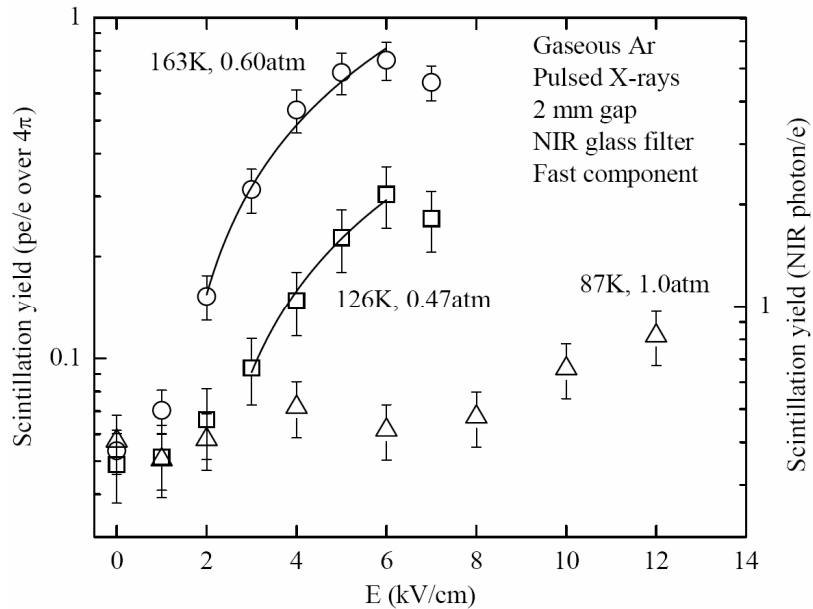

Fig. 4. Primary and secondary scintillation yield in the NIR for the fast emission component as a function of the electric field in gaseous Ar, at different cryogenic temperatures and pressures: at 163 K and 0.60 atm, at 126 K and 0.47 atm, at 87 K and 1.0 atm. The scintillation yield is given in the number of GAPD photoelectrons per ionization electron (left scale) and in the number of NIR photons per ionization electron (right scale), both quantities normalized to the full solid angle ($4\pi$). The data were obtained under X-ray irradiation in a 2 mm thick gap (top gap). The curves represent linear fits of the data in the proportional scintillation mode.

Accordingly, the primary scintillation yield in gaseous Ar, defined by formula (1), was measured at lower electric fields, i.e. in the absence of secondary scintillations, in the range of 0-1 kV/cm: it is shown in Fig. 3. Three groups of data are presented obtained at similar cryogenic temperature, of about 160 K, but under substantially different conditions: with the top GAPD and NIR glass filter, with the top GAPD and borosilicate glass filter, with the bottom GAPD and no optical filter. One can see that there is a good agreement between all the data presented within their accuracies.

Hence one may conclude the following. First of all, the scintillations took place essentially in the NIR, since the light yields obtained with and without the NIR filter were close to each other. Moreover, the NIR emission spectrum in our measurements was similar to that of ref. [28], since just this spectrum provides the data compatibility. Furthermore, the acceptances were calculated correctly, since the data obtained in the top and bottom gaps are consistent, despite the considerable difference in measurement geometries and GAPD solid angles (by more than an order of magnitude [36]).

Thus, the primary scintillation yield in the NIR in gaseous Ar, averaged over all the measurements presented in Fig. 3, amounts to

$$Y_{GAr} = 0.44 \pm 0.09 \text{ photon/e} = 17000 \pm 3000 \text{ photon/MeV} . \qquad (6)$$

It is convenient to express here the yield in the number of photons per MeV of deposited energy considering a W-value (energy needed to produce one ion pair) in gaseous Ar of 26.4 eV [3].

It should be remarked that the primary scintillation yield in the NIR was found to be independent of temperature in the range of 87-163 K. This is seen from Fig. 4 showing the primary scintillation (below 1 kV/cm) and secondary scintillation (above 2 kV/cm) yields for the fast emission component as a function of the electric field, in gaseous Ar at different cryogenic temperatures and pressures. It is also interesting that the primary scintillation yield in the NIR is close to that of the VUV for gaseous Xe [24] and to that of the low limit established for NIR scintillations in Xe [34]. Thus, one may conclude that the primary scintillation yield is rather similar for all noble gases, both in the VUV and NIR regions.

## 4. Electroluminescence yield in gaseous Ar

From Figs. 2 and 4 one can see that the secondary scintillation yield in gaseous Ar increased with the electric field starting from a certain threshold: in particular at 163 K - from 1.5 kV/cm and at 126 K - from 2 kV/cm, at a pressure of 0.60 and 0.47 atm respectively, i.e. at the gas density corresponding to that of 1 atm and room temperature. This increase is explained by secondary scintillations (electroluminescence) induced by electrons drifting and exciting atoms under moderate electric fields, at which the gap is operated in an ionization mode, i.e. without electron multiplication. As considered in section 2, this kind of scintillations is called proportional scintillations if their intensity is proportional to the electric field in a certain field range. In denser gaseous Ar, at 87 K and 1 atm, electroluminescence was also observed at fields exceeding 10 kV/cm (see two last points in Figs. 4), albeit on the verge of its appearance.

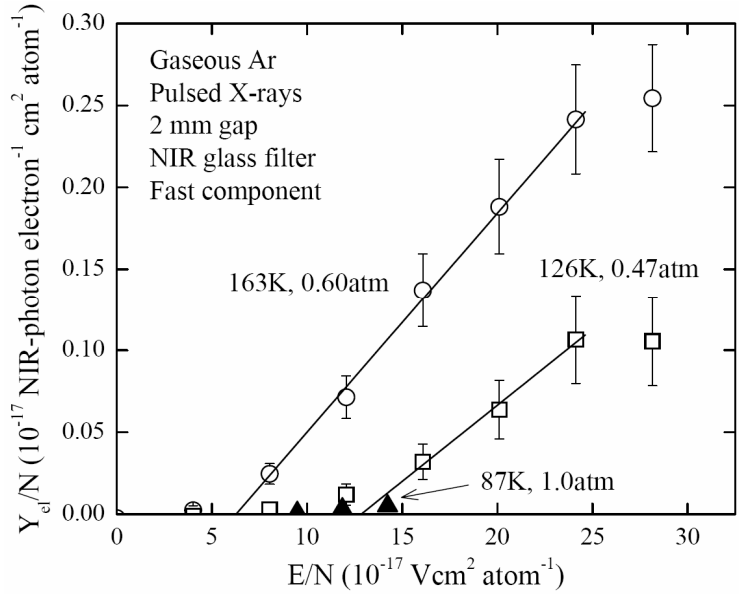

Fig. 5. Reduced NIR electroluminescence yield for the fast emission component as a function of the reduced electric field in gaseous Ar at 163 K and 0.60 atm, 126 K and 0.47 atm, 87 K and 1.0 atm, measured in the top gap with NIR glass filter in front of the GAPD.

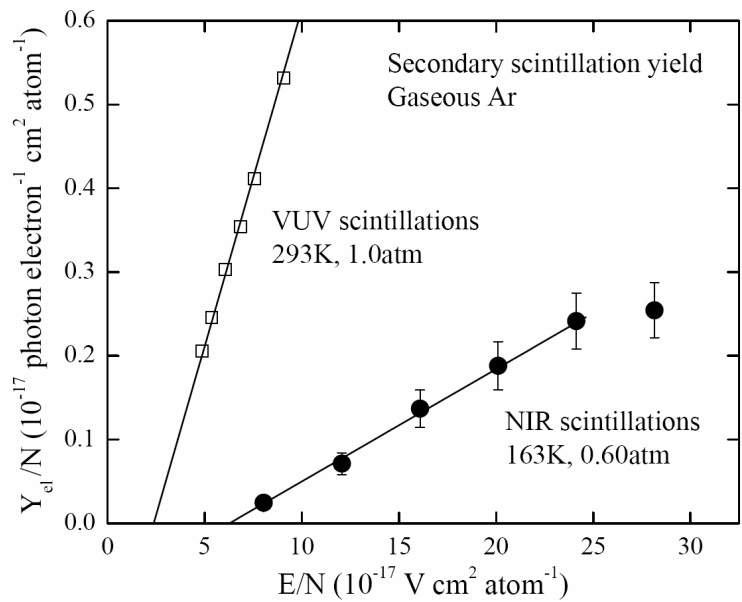

Fig. 6. Comparison of the reduced electroluminescence yield in gaseous Ar in the NIR measured in the present work, for the fast emission component at 163 K and 0.60 atm, to that in the VUV measured elsewhere [25].

The effect of proportionality of NIR scintillations in Ar was observed earlier at room temperature [30], though the absolute scintillation yield was not measured that time. In the present work, in gaseous Ar at 163 K and 126 K, the law of proportionality holds up to 6 kV/cm. At higher fields, the electron multiplication started (see Fig. 2), resulting in saturation of the electroluminescence yield. Nevertheless at this field, the scintillation yield increased by an order of magnitude compared to that of primary scintillations, reaching a value of 5 photon/e (or $200 \times 10^3$ photon/MeV) for the average electron drift path of 1 mm: see Fig. 4.

Consequently, we confirm here a possibility to reach rather high yields for secondary scintillations in the NIR, similarly to that observed recently in our laboratory for secondary avalanche scintillations using THGEM multipliers [2],[21]: the avalanche scintillation yield of 4 NIR photon per avalanche electron was observed there at charge gains of 60-400.

In fact, the right variables to study proportional scintillations are those defined by equations (2)-(5) in section 2, i.e. the reduced electroluminescence yield and the reduced electric field: these are shown in Fig. 5. One can see again that proportional scintillations exist in the NIR up to reduced field values of 25 Td and that at higher fields the electroluminescence yield is saturated due to the commencement of the electron multiplication. It seems that the slope of the field dependence of the electroluminescence yield, which characterizes the proportional scintillation efficiency, weakly depends on the temperature.

On the other hand, the increase of the electroluminescence threshold with the temperature decrease was observed (see Fig. 5). In particular in Ar at 87 and 1 atm the threshold was so high that, despite of the onset of electroluminescence, the proportional scintillation mode was not reached in the field range studied (up to 12 kV/cm). At the moment we cannot explain the dependence of the electroluminescence threshold on temperature. It should be noted however that likely the similar behavior of the electroluminescence yield in the VUV range, namely its possible dependence on temperature, was discussed in ref. [25].

To compare our electroluminescence results obtained in the NIR to those obtained in the VUV, the reduced electroluminescence yield is shown in Fig. 6 as a function of the reduced electric field. The linear part of the field dependence corresponds to proportional scintillations in the NIR; at 163 K it is described by the following equation:

$$\frac{Y_{el}}{N} [10^{-17} \text{ photon electron}^{-1} \text{ cm}^2 \text{ atom}^{-1}] = 0.013 \frac{E}{N} - 0.084 . \qquad (7)$$

Here $E/N$ is given in Td [$10^{-17}$ V cm$^2$ atom$^{-1}$]. This equation may be compared to the proportional scintillation yield in the VUV at room temperature presented elsewhere [25]:

$$\frac{Y_{el}}{N} [10^{-17} \text{ photon electron}^{-1} \text{ cm}^2 \text{ atom}^{-1}] = 0.081 \frac{E}{N} - 0.190 . \qquad (8)$$

One can see that the electroluminescence amplification parameter in the NIR, defined as the number of photons produced per drifting electron and per kilovolt, equivalent to the line slope in Fig. 6, is equal to 13; this is by a factor of 6 lower compared to that of the VUV. That means that proportional scintillations in the NIR are substantially less efficient than in the VUV.

It should be remarked that recent simulations of the Ar electroluminescence yield in the NIR were in fair agreement with our experimental data [40]. Finally, though having somewhat lower yield than that in the VUV, proportional scintillations may considerably increase the scintillation yield in the NIR as compared to that of primary scintillations: by an order of magnitude, to hundreds of thousands photons per MeV.

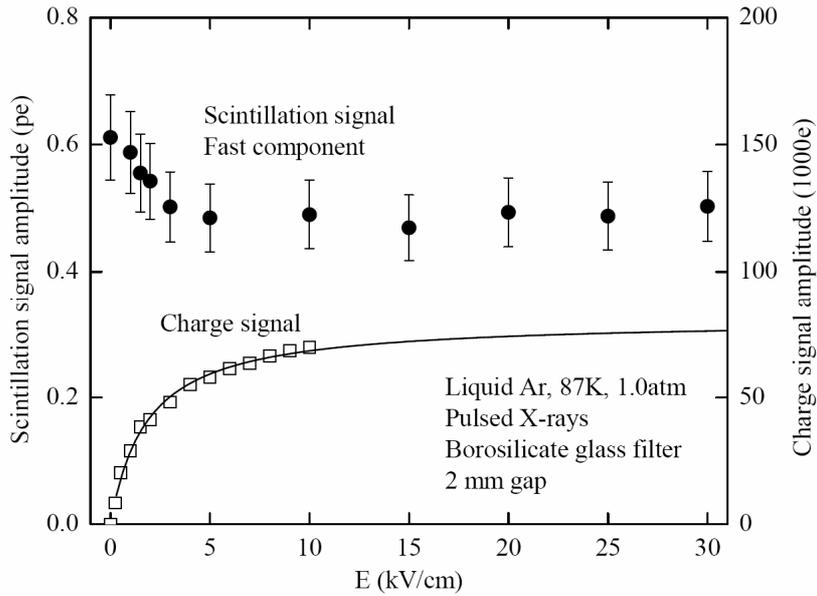

Fig. 7. Typical scintillation and charge signals amplitudes, measured in the top gap in liquid Ar at 87 K and 1.0 atm, with borosilicate glass filter in front of the GAPD. Shown are the scintillation signal amplitude of the fast emission component expressed in photoelectrons (left scale) and the charge signal amplitude expressed in electrons (right scale), per X-ray pulse, as a function of the electric field in the gap. The curve is the recombination model fit to the data points.

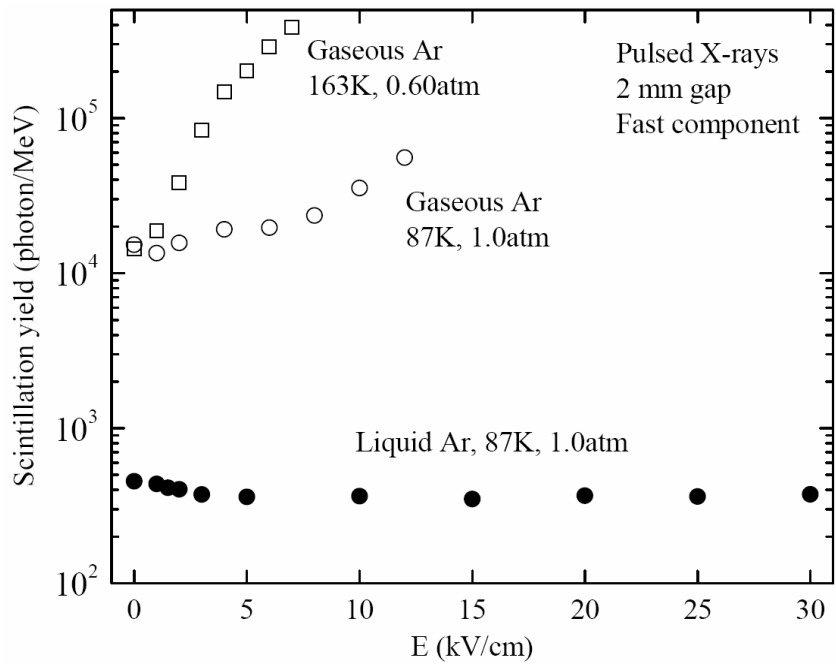

Fig. 8. Scintillation yield for the fast emission component as a function of the electric field in liquid Ar in the NIR and visible region, at 87 K and 1.0 atm, and in gaseous Ar in the NIR, at 163 K and 0.60 atm and at 87 K and 1.0 atm. The scintillation yield is given in the number of photons per MeV of deposited energy of the primary ionization. The data were obtained under X-ray irradiation in a 2 mm thick gap (top gap).

## 5. Primary scintillation yield in liquid Ar

In liquid Ar, the charge collected in the gap ($N_c$) strongly depends on the electric field ($E$) due to the recombination effect: see Fig. 7 showing typical scintillation and charge signals amplitudes measured in the top gap in liquid Ar at 87 K and 1.0 atm. In our measurements this dependence was described in the frame of the recombination model presented in ref. [4], enabling to determine the primary ionization charge ($N_e$) from the following formula:

$$N_c = \frac{N_e}{1 + k/E} \ . \tag{9}$$

Here $k$ is the fitting parameter, depending on the primary ionization density. In particular the curve in Fig. 7 is the result of such fitting to the experimental data.

In liquid Ar the non-VUV emission spectrum was reported to be continuous [31],[32] extending from 400 to 1000 nm, in contrast to the atomic line spectrum in gaseous Ar. Accordingly, in liquid Ar we used the borosilicate glass filter transparent in both the NIR and visible range. Taking these facts into account, Fig. 8 shows the scintillation yield in liquid Ar at 87 K, in the NIR and visible range, as a function of the electric field varying from 0 to 30 kV/cm; the yield was calculated using formula (1). For comparison, the NIR scintillation yield in gaseous Ar is also shown at the same temperature, i.e. in saturated vapor at 87 K and 1.0 atm, as well as at 163 K and 0.60 atm. The scintillation yield in Fig. 8 is given in the number of photons per MeV of deposited energy of the primary ionization, using a W-value of 26.4 and 23.6 eV for gaseous and liquid Ar respectively [3]. The primary scintillation yield in liquid Ar in the NIR and visible range, defined at zero field, amounts to

$$Y_{LAr} = 510 \pm 90 \ \text{photon/MeV} \ . \tag{10}$$

Note that it is considerably reduced, by a factor of 40, compared to that of gaseous Ar. The effect of suppression of NIR scintillations in the noble liquid as compared to that of the noble gas was observed earlier for Xe [29]. The mechanism of such suppression is yet unclear.

Finally, in liquid Ar in contrast to gaseous Ar, no electroluminescence was observed at all, even at fields reaching 30 kV/cm: see Fig. 8. It should be remarked however that we did not reach here the characteristic field value of 60 kV/cm which was reported to be the threshold for VUV electroluminescence, observed in THGEM and GEM plates immersed in liquid Ar [2],[20].

## 6. Possible applications of NIR scintillations

The possible applications of NIR scintillations have been already discussed in ref. [2]. These included two-phase Ar CRADs with THGEM/GAPD optical readout for rare-event experiments, high-energy noble-liquid calorimetry with optical readout in the NIR and imaging detectors using CCD-based optical readout in the NIR, including those of digital radiography. In the present paper we further discuss the applications of NIR scintillations, namely the elaborated project of the two-phase Ar CRAD with THGEM/GAPD optical readout for rare-event experiments and the NIR scintillation monitor for ion beam radiotherapy.

### 6.1 Two-phase Ar CRAD with THGEM/GAPD optical readout for rare-event experiments

At the moment, the project on the development of the two-phase Ar CRAD with THGEM/GAPD optical readout in the NIR is being carried out in our laboratory in the frame of the Grant of the Government of Russian Federation (11.G34.31.0047). The final goal of the project is the development of the detector of ultimate sensitivity for dark matter search and coherent neutrino-nucleus scattering experiments. The detector should be capable to count single electrons of primary ionization produced in the liquid at a very low deposited energy (<1 keV), at high spatial resolution (<1cm) and extremely low noise. The principles of operation of such a CRAD have been presented elsewhere [1],[2]. Fig. 9 shows the elaborated version of the detector; its principle features are proposed as follows. The detector consists of a vacuum-insulated cryogenic chamber of a 50 cm diameter and 100 cm height, having a volume of 200 l. The maximum drift length in the liquid of 50 cm and the active area of a diameter of 36 cm define the fiducial volume of the detector of 50 l corresponding to 70 kg of liquid Ar. The total amount of liquid Ar in the chamber is 150 l (200 kg). The low radioactivity Ar from underground sources [41] will be used, to minimize the background due to $^{39}$Ar isotope.

The electrons of primary ionization, produced by nuclear recoil from a weakly interacting particle, drift in the liquid towards its surface under an electric field of 2 kV/cm. We also consider a possibility to increase the drift field by a factor of 2 in order to increase the charge yield from a track, i.e. to reduce the recombination effect in liquid Ar [42]. The electrons are emitted from the liquid into the gas phase, where they are amplified in two stages: first, using proportional scintillations in the VUV produced in an electroluminescence gap (EL gap) above the liquid-gas interface and, second, using avalanche scintillations in the NIR produced in a double-THGEM multiplier or in a triple structure like double-THGEM/GEM.

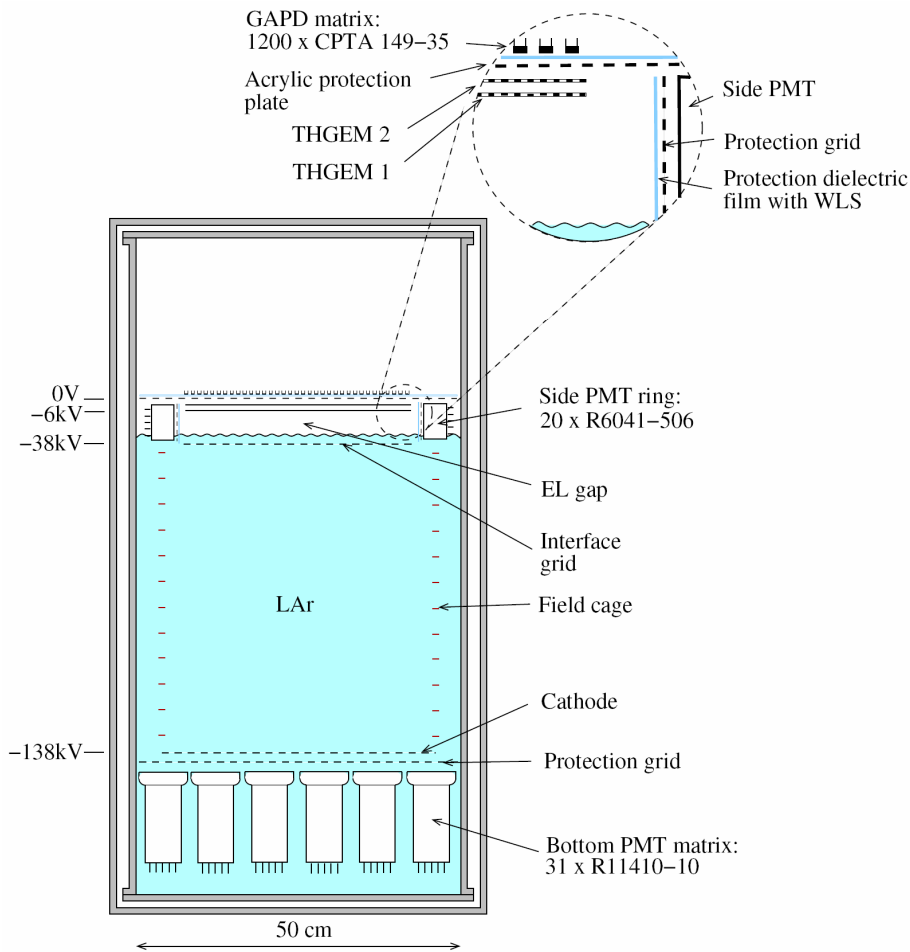

Fig. 9. Schematic view of the two-phase Ar CRAD with THGEM/GAPD optical readout in the NIR for dark matter search and coherent neutrino-nucleus scattering experiments (to scale). This project is being developed in our laboratory.

Proportional scintillations are recorded in the VUV using two PMT arrangements: the bottom PMT matrix placed in the liquid at the chamber bottom and the side PMT ring placed around the EL gap. The bottom PMT matrix is composed of 31 low-background cryogenic 3-inch PMTs, namely of Hamamatsu R11410-10 or R11065-10 [43],[44],[45] with bialkali photocathode (64 mm in diameter) and box-and-linear-focused dynode system. The side PMT ring is composed of 20 low-background cryogenic 2-inch PMTs, namely of Hamamtsu R6041-506 [43] with bialkali photocathode (45 mm in diameter) and metal-channel dynode system, the latter making them rather compact. To be sensitive to VUV scintillations, all the PMTs are supplied with a WLS coating (Tetraphenyl-Butadiene, TPB [46],[47]), deposited either on a transparent dielectric film in front of PMTs or directly on them.

The bottom PMTs will be also used for the detection of primary scintillations produced in the fiducial volume.

The bottom and side PMTs provide a single- and double-electron trigger for primary ionization due to the excellent amplitude resolution available in the proportional scintillation mode. The reduced solid angle ($\Delta\Omega/4\pi$) of the bottom PMT matrix and that of the side PMT ring with respect to the EL gap is 3% and 6% respectively. To avoid electrical discharges from the HV electrodes to the PMTs and field penetration into the dynode system, the bottom PMTs are protected with a wire grid, in addition to the liquid Ar layer acting as a good insulator, while the side PMTs are protected with a transparent dielectric film (coated with the WLS) and another wire grid.

The EL gap, having a thickness of 4 cm, matches with the size of the side PMT. In addition, the electric field in the gap, of 8 kV/cm corresponding to the voltage of 32 kV applied across the gap, is quite high to emit enough light for providing the single-electron sensitivity. Moreover, such a high electric field is a good thing for efficient electron extraction from liquid Ar [5],[48]. According to Fig. 6, the number of photons emitted in the VUV per drifting electron at this field (E/N=10 Td) will be 2000. Taking into account the overall reduced solid angle of the side and bottom PMTs (9%), their quantum efficiency (25%) and WLS reemitting efficiency (50%), the total number of photoelectrons recorded by both PMT arrangements will be 23 pe. This is enough to make a selection between single- and double-electron events.

Avalanche scintillations produced in the holes of the second THGEM are recorded in the NIR using a matrix of GAPDs: this will provide a high (sub-cm) spatial resolution. The matrix is composed of 1200 GAPDs of a CPTA 149-35 type [37], having a 2.1×2.1 mm$^2$ active area each, and placed with a pitch of 1 cm in a hexagonal-type lattice, at a distance of 1 cm from the second THGEM. Such an optical readout is preferable as compared to charge readout in terms of overall gain and noise. In particular it has been demonstrated [21] that the combined THGEM/GAPD multiplier could operate in a single electron counting mode in the two-phase Ar CRAD, at charge gains exceeding 500. It has been recently confirmed that the double-THGEM multiplier of a 10×10 cm$^2$ active area can provide such gains in two-phase Ar CRADs [2],[49]. Even higher gains, by an order of magnitude, can be obtained in the triple-structure studied recently in our laboratory [49], namely in that of double-THGEM/GEM.

The use of the side PMTs, with their solid angle exceeding that of the bottom PMTs, in principle admits to do without the bottom PMTs when detecting proportional scintillations. In this case the fiducial volume can be further increased by increasing the maximum drift length. In addition, the interesting technical solution for the EL gap might be realized in this case: the interface grid in the liquid might be replaced with the more robust THGEM plate, effectively transmitting the drifting electrons through its holes under a certain voltage applied between its electrodes. Such a design will help to avoid the flatness problem particularly important for the EL gap operated at such high electric fields.

An alternative (but less preferred) solution is to give up of the side PMTs. In this case the bottom PMTs should be placed closer to the EL gap, to regain the solid angle. The decrease of the fiducial volume can be partly compensated by the increase of the active area, which was difficult to obtain in the previous design due to the side PMTs. In particular, if to take the active area of a diameter of 45 cm and the maximum drift length of 20 cm, the fiducial volume will amount to 30 l, which is not that small. At the same time the reduced solid angle of the bottom PMTs will be of 10%, which is enough for single electron treatment.

The distinctive feature of the detector concept (first suggested in [12]) is the selection of point-like events having 2 or more primary ionization electrons, to reject the single-electron background. Accordingly, compared to other noble liquid detectors for rare-event experiments, the proposed detector will have the advantage of the higher spatial resolution at a lower detection threshold, which will increase its sensitivity in dark matter search and coherent neutrino-nucleus scattering experiments.

## 6.2 NIR scintillation monitor for ion beam radiotherapy

The second application considered in the present work is the NIR scintillation position monitor for ion beam radiotherapy. In the last decade, ion beam radiotherapy has become an emerging technique used for the treatment of solid cancers [50],[51],[52]. On the other hand, the space selectivity of this therapy asks for new approaches to the delivered dose monitoring, since a rather precise position monitoring of the dose is essential for a good quality control of the treatment. In particular the required position resolution should be of sub-mm level.

It should be remarked that for the last few years the project of "Heavy Ion Therapy System" has been under development at Budker INP [52]. The presence of electron cooling in the ion synchrotron provides a small size and small energy spread of the cooled beam, thus enabling the realization of the novel beam extraction scheme by small precisely-dosed portions, the so-called "pellet" extraction. Due to the high dose rate in one "pellet", the traditional approaches for beam monitoring, for example based on gaseous ionization chambers, are not applicable. Here we suggest a new approach to such monitoring: we propose the monitor of the ion beam position based on recording NIR scintillations in gaseous Ar.

The idea of the monitor consists of recording primary NIR scintillations in gaseous Ar induced by the ion beam, using an arrangement of NIR-sensitive PMTs as shown in Fig. 10. This arrangement is the ring of 20 PMTs placed in a circle of a diameter of 34 cm around the beam. Such a diameter was chosen to encompass the beam displacement in cross direction limited to 20 cm. The NIR scintillation amplitudes are measured with the PMTs. To calculate the ion beam position, a dedicated algorithm was developed based on comparison of these amplitudes.

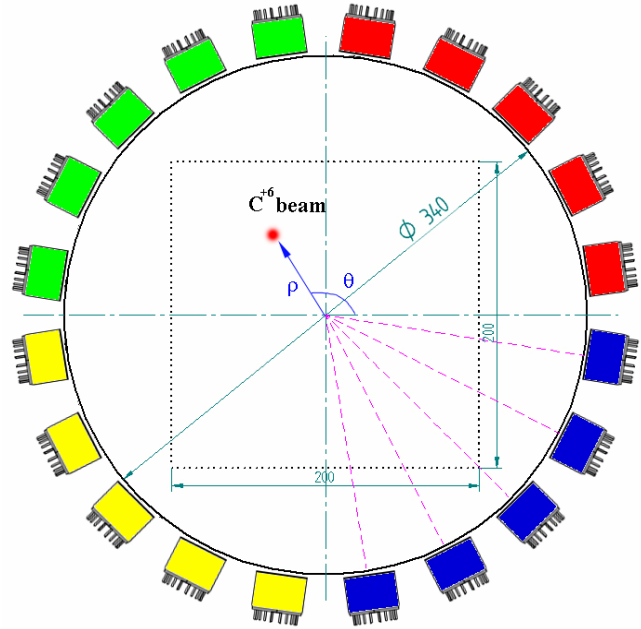

Fig. 10. Schematic view of the NIR scintillation position monitor for ion beam radiotherapy, composed of 20 PMTs placed in a circle of a diameter of 340 mm around the beam, with either multi-alkali or GaAs photocathode of a diameter of 18 mm, recording primary scintillations in the NIR induced by the Carbon ion beam. The monitor has a thickness of 2 cm and is filled with Ar at 1 atm. The limits of the beam displacement in cross direction are also shown.

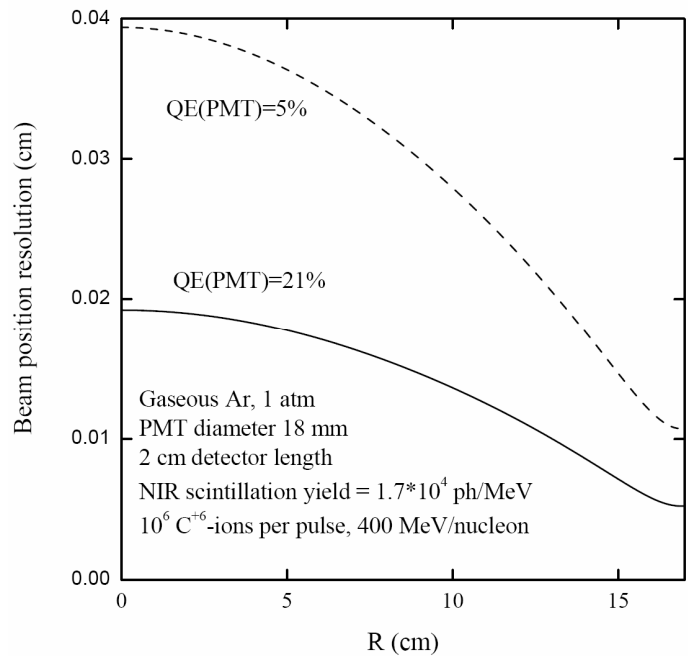

Fig. 11. Beam position resolution as a function of the beam distance with respect to the center of the monitor shown in Fig. 10, for PMTs with multi-alkali and GaAs photocathode having the quantum efficiency of 5% and 21% respectively, averaged over the gaseous Ar emission spectrum. The beam parameters are as follows: $10^6$ Carbon ions per pulse ("pellet") with energy of 400 Mev/nucleon, depositing 130 keV/cm in gaseous Ar.

For the particular calculations, we took the characteristics of the existing PMTs manufactured by the Ekran company [53]. These PMTs have a MCP-type dynode system and either multi-alkali or GaAs photocathode, of a diameter of 18 mm, having the quantum efficiency of 5 and 21 % respectively [38] averaged over the NIR emission spectrum of gaseous Ar. The monitor has a thickness of 2 cm and is filled with Ar at 1 atm. The primary scintillation yield of $1.7 \times 10^4$ photon/Mev was taken in the calculations.

The results of the calculations of the position resolution for the typical Carbon ion beam used in radiotherapy ($10^6$ Carbon ions per "pellet" with energy of 400 Mev/nucleon) are presented in Fig. 11. One can see that the resolution is improved when using the PMTs with GaAs photocathode and when the beam is displaced to the periphery. But even at the worst, when the beam is in the center and the PMTs with multi-alkali photocathode are used, the position resolution is rather high, of 0.4 mm. This proves the concept principle.

## 7. Conclusions

Following the first results obtained in our laboratory [1] and preceded by part I [36], here in Part II we presented a comprehensive study of the light yield of primary and secondary scintillations produced in gaseous and liquid Ar in the NIR and visible region at cryogenic temperatures. The measurements were performed using Geiger-mode avalanche photodiodes (GAPDs) and pulsed X-ray irradiation; the appropriate methodology was considered in Part I.

In gaseous Ar, the non-VUV scintillations took place essentially in the NIR. The primary scintillation yield of the fast emission component in gaseous Ar was found to be independent of temperature in the range of 87-160 K; it amounted to $17000 \pm 3000$ photon/MeV in the range of 690-1000 nm. This is comparable with that of the VUV for gaseous Ar and Xe.

In liquid Ar, the primary scintillation yield of the fast component was considerably reduced, amounting to $510 \pm 90$ photon/MeV, in the range of 400-1000 nm. Nevertheless this is comparable with the yields of fast solid scintillators used in high-energy calorimetry.

Proportional NIR scintillations (electroluminescence) in gaseous Ar were also studied; their amplification parameter at 160 K was measured to be 13 photons per drifting electron per kV. The possible increase of the electroluminescence threshold with the temperature decrease needs to be confirmed. Though having somewhat lower yield than that in the VUV, proportional scintillations may substantially increase the scintillation yield as compared to that of primary scintillations in the NIR: by an order of magnitude, to hundreds of thousands photons per MeV.

No proportional scintillations were observed in liquid Ar up to the electric fields of 30 kV/cm.

There is potentially a wide variety of applications of noble gas NIR scintillations in high energy physics experiments and medical applications [2]. In the present work we proposed the elaborated version of the two-phase Cryogenic Avalanche Detector (CRAD) in Ar with THGEM/GAPD optical readout in the NIR, for dark matter search and coherent neutrino-nucleus scattering experiments. Such a detector will have the advantage of the higher spatial resolution at a lower detection threshold. In the medical application field, we proposed the NIR scintillation position monitor for Carbon ion beam radiotherapy.

Further studies in these directions are in progress in our laboratory.

## 8. Acknowledgements

We are grateful to A. Chegodaev and R. Snopkov for technical support, M. Barnyakov and E. Kravchenko for collaboration and D. Akimov and Y. Tikhonov for discussions. This work was supported in part by the Ministry of Education and Science of Russian Federation and Grant of the Government of Russian Federation (11.G34.31.0047).

## Appendix

In a parallel-plate gap having the thickness $d$, operated in an electron-avalanching mode with the ionization coefficient $\alpha$ and charge gain $G$, we have the equation

$$G = \exp(\alpha\, d)$$

Consequently, the total drift path of all the electrons produced in such a gap is

$$L = \int_0^d \exp(\alpha\, x)\, dx = \frac{1}{\alpha}\big[\exp(\alpha\, d) - 1\big] = d\,\frac{G-1}{\ln G}$$

From here one can easily derive equation (4) for the average drift path per electron, taking into account that the number of electrons is equal to $G$ and that the primary ionization is produced uniformly across the gap.